# Resolving ultrahigh–contrast ultrashort pulses with single–shot cross–correlator at the photon noise limit

Jingui Ma,[1] Peng Yuan,[1] Xiaoping Ouyang,[2] Jing Wang,[1] Guoqiang Xie,[1] and Liejia Qian[1,*]

[1] Key Laboratory for Laser Plasmas (MOE), School of Physics and Astronomy, and Tsung-Dao Lee Institute, Shanghai Jiao Tong University, Shanghai 200240, China

[2] National Laboratory on High Power Laser and Physics, Shanghai Institute of Optics and Fine Mechanics, Chinese Academy of Sciences, Shanghai 201800, China

*email: qianlj19@sjtu.edu.cn

**Abstract:** In strong-field physics experiments with intense lasers, it is of paramount importance to single-shot diagnose the temporal contrast between laser pulse peak and its noise pedestal. This allows fast optimization of pulse contrast and meaningful comparison with theory for each pulse shot, and it can help new outcomes from clean laser-plasma interactions. Thus far, high contrast ratios up to ~$10^{10}$, required by present petawatt (PW) class lasers, have been accessible in both generation and single-shot characterization. However, ultrahigh contrast ~$10^{13}$, required by the planned 200-PW lasers, challenges intense laser technology and remains an open question. This paper reports on the first demonstration of such an ultrahigh-contrast measurement by adapting single-shot cross-correlator (SSCC). We introduce an ultrafast method that enables to determine the SSCC detection limit. Our strategy mimics the test laser having known ultrahigh contrast in the measurement frame of time-to-space mapping. The ultimate contrast-measurement limit of $10^{13}$ is achieved, which corresponds to the highest pulse intensity set by SSCC damage threshold and the lowest noise pedestal set by single-photon detection. As a consequence, photon noise in the detection is observed and increases as the noise pedestal reduces. The demonstrated measurement ability at the photon noise limit is applied to a high-contrast laser system based on second-harmonic generation and optical parametric chirped-pulse amplification, suggesting accessible of ultrahigh contrast pulses.

Since the invention of chirped-pulse amplification, femtosecond lasers can now deliver ultrahigh peak powers beyond petawatt (PW), paving the way toward compact particle accelerators and X-ray sources [1]. One crucial parameter for strong-field physics experiments is the temporal intensity contrast between the laser pulse peak and its noise pedestal, which should be large enough to avoid an early ionization of the plasma target before the laser pulse arrives [2-4]. Consequently, present PW-class lasers with $\sim10^{21}$ W/cm$^2$ intensity should be matched with a high contrast of $10^{10}$, while the contrast requirement will climb to $\sim10^{13}$ for the planned 200-PW lasers (e.g., the Extreme Light Infrastructure) aiming at an extreme intensity of $10^{24}$ W/cm$^2$ [5]. Such demanding contrast of $10^{13}$ is approximately 5 to 6 orders of magnitude better than those of femtosecond mode-locking lasers [6].

Although femtosecond pulse measurements have been well developed, ultrafast diagnosis of pulse contrast with ultrahigh dynamic-range imposes daunting technological challenges. The state-of-the-art delay-scanning cross-correlator (DSCC) can provide a high dynamic-range of $10^{12}$ to $10^{13}$, but scanning measurement is heavily time consumed and only suitable for repetitive lasers. For example, the front-end measurement of Apollon 10-PW laser indicated a dynamic-range-limited contrast as high as $10^{12}$ [7]. However, final output contrast of an ultrahigh power laser is usually inaccessible because of low repetition rate. On the other hand, DSCC measurement is an average result of repetitive detections, which may deviate from real pulse contrast in some degree because the laser noise varies from pulse shot to shot. Single-shot cross-correlator (SSCC) becomes essential in diagnosing and optimizing final contrast of a laser system [8]. Thus far, SSCC has been successfully applied to several PW-class lasers with $\sim10^{10}$ contrast [9, 10], which is the result of a long series of improvements over the last decades [11-18]. It is highly desired that SSCC measurement can resolve ultrahigh contrast of $10^{13}$ for the coming 200-PW lasers.

Here we demonstrate an ultrafast method to evaluate the contrast measurement ability of an SSCC and highlight the ultimate measurement limit set by detection photon noise. While the dynamic range of DSCC measurement can easily be tested by optical attenuation adjusted during its long scanning time, it is assumed that the dynamic-range test of SSCC measurement must rely on an ultrashort pulse with a known contrast better than $10^{13}$. In addition, the test pulse must have high enough power, e.g., ~100-GW power to achieve high intensity within the correlation area of ~10 × 10 mm$^2$. Such a required high-power ultrahigh-contrast pulse has not yet been available, and hence all the SSCC designs have not been tested in practice. In this work, we mimic a test pulse in the spatial domain based on time-to-space mapping, and evaluate the efficacy of SSCC device to measure $10^{13}$ contrast.

The principle behind SSCC device is simple: A femtosecond test pulse with its picosecond noise pedestal is recorded by oscilloscope with an adaptor. Here the adaptor plays a key role of temporal magnification, and it converts the test pulse into a series of nanosecond spaced temporal slices. As shown in Fig. 1a, the adaptor relies on a correlation unit of noncollinear third-harmonic generation (THG) and a detection unit of fiber-array-mediated photomultiplier tube (PMT) (see the detail in Methods). The noncollinear THG performs time-to-space mapping that enables a test pulse $I(t)$ represented by its correlation function $A^{(2)}(x)$ in the spatial domain (Fig. 1a). With the help of a 100-pixel fiber array, $A^{(2)}(x)$ is further mapped into a series of temporal slices spanning 500 ns. Thus such a temporally magnified signal series can be resolved by PMT and oscilloscope. This is the routine procedure of single-shot contrast measurement. But, a spatially shaped beam $I(x)$ can also give rise to the same $A^{(2)}(x)$ as the test pulse $I(t)$ does (Fig. 1b). This suggests an alternative for the test of device dynamic-range. As shown in Fig. 1c, a test beam $I(x)$ is synthesized by spatially packaging a high-intensity narrow beam in the center of a surrounding continuous wave (CW) wide beam. Under the measurement frame of time-to-space mapping, the femtosecond intense beamlet mimics a peak pulse under test, and the CW weak beamlet mimics a temporal noise pedestal. In other words, the synthesized beam acts equivalently as a test pulse having a known contrast set by the intensity ratio of two beamlets.

In the test experiment for 1054-nm SSCC, the correlation unit adopted a $\beta$-BBO crystal with a size of 38 (x) × 10 (y) × 2 (thickness) mm$^3$ and a noncollinear angle $\alpha$=13.8°, which induces a mapping coefficient 13.3 ps/cm and a fiber-pixel time 500 fs. The 1-kHz repetition rate femtosecond beamlet with a width $W_x$=0.38 mm was obtained by second-harmonic generation (SHG) of a 2108-nm optical parametric amplifier (OPA) system and had a fixed intensity of 30 GW/cm$^2$ at the crystal surface, while the CW beamlet with $W_x$=38 mm had a varied intensity. Since the high-intensity (~60 GW/cm$^2$) sampling beam at 527 nm in our SSCC device also had a narrow width $W_x$=0.38 mm, the test experiment was performed by beam scanning along the x direction. As shown in Fig. 2a, the measured contrast by our device agrees well with the intensity ratio of two beamlets. The detectable CW beamlet intensity can be as low as 3.5 mW/cm$^2$, indicating an ultrahigh dynamic-range of ~10$^{13}$.

The recorded average photovoltage of 3.1 mV at the CW intensity of 3.5 mW/cm$^2$ corresponds to only one THG photon from a fiber pixel (Figs. 2b and 2d), inferred by the PMT parameters (see Methods for details). In this situation, the shot-to-shot fluctuations are on the order of 250%. Such behavior is typical for single-photon detection with the characteristics of white noise. At higher THG photon number (>10), the fluctuations get reduced greatly (Fig. 2c), and the distribution of detected photon numbers differs slightly from a Poisson distribution with the same average flux (Fig. 2e). Consequently, the observed severe detection fluctuations due to photon noise will eventually limit SSCC measurement for ultrahigh contrast pulses. At this point, it is also clear that DSCC measurement may not be reliable owing to these fluctuations of photon noise and laser noise. The ultimate measurement limit of 10$^{13}$ contrast achieved here nearly corresponds to the highest pulse intensity set by SSCC damage threshold and the lowest noise pedestal set by single-photon detection. Notably, the demonstrated measurement ability can be linearly scaled up with the transverse size ($W_y$) of test laser beam and/or correlation crystal. Since nonlinear crystals such as LBO and YCOB can be made as large as $W_y \approx$100 mm [19], we anticipate an attainable extreme dynamic-range of ~10$^{14}$.

Figure 3 presents two SSCC measurements for a real ultrashort pulse. The test laser was achieved from SHG of a single-shot optical parametric chirped-pulse amplification (OPCPA) system at 2108 nm (see Methods for details). The 2108-nm femtosecond OPA laser served as the clean seeder to OPCPA with a 22-mm-thick LiNbO$_3$ crystal and a 1054-nm pump laser with intensity of 1 GW/cm$^2$. The test laser system generated ~100-GW ultrashort pulses at 2108 nm and ~40-GW ultrashort pulses at 1054 nm. In the SSCC measurements at 1054 nm, the temporal window was increased to 120 ps by using a large noncollinear angle $\alpha$=33.6°.

SHG of the OPCPA output can produce ultrahigh contrast pulses. However, the full-window (120 ps) single-shot measurement by SSCC shows a ratio of only ~$3\times10^{10}$ around -105 ps, limited by the low intensity of total test pulse (10 GW/cm$^2$). Real contrast as high as ~$0.5\times10^{13}$ can be confirmed by a small-window (24 ps) measurement with high test pulse intensity (50 GW/cm$^2$), as shown in the six-shot plot that consists of six equally-delayed (20 ps) single-shot measurements. The improved contrast measurement over two orders of magnidude consists well with the theoretical expectation that THG correlation signal has a cubic dependence with the total intensity of test pulse. It is worth mentioning that a real single-shot measurement of $10^{13}$ contrast will be achieved by using a 200-GW test pulse at 1054 nm.

In conclusion, we have demonstrated the photon-noise-limited single-shot measurement of ultrahigh contrast. To the best of our knowledge, this $10^{13}$ contrast measurement represents the highest reported dynamic-range to date in single-shot ultrafast diagnostics. The demonstrated ability is ready for a broad diagnosing from current PW lasers to the coming 200-PW lasers and beyond.

## Methods

**Light sources.** As shown in Fig. 4, two kinds of test laser were applied in the experimental study. One was a high spatial-contrast beam consisting of high-intensity femtosecond narrow beam and low-intensity CW wide beam, and the other was a high temporal-contrast pulse from SHG of a single-shot OPCPA laser. Each test laser can be independently coupled into SSCC device by a translation stage TS-1. An 1-kHz femtosecond OPA (OPerA SOLO, Coherent) pumped by Ti:sapphire regenerative amplifier (Astrella, Coherent) delievered 330-μJ, 50-fs pulses at 2108 nm, which was further converted to 120-μJ femtosecond pulses at 1054 nm by SHG with an 1-mm-thick LiNbO$_3$ crystal. This 1-kHz, 1054-nm femtosecond laser, and a single-frequency CW fiber laser (Rock single-frequency laser, NP Photonics), were synthesized to a high-contrast beam that was the test laser in Fig. 2. Hereinafter, the second test laser source will be described, and it relied on SHG of a 2108-nm optical parametric chirped-pulse amplification (OPCPA) system. The OPCPA pump laser consisted of the above mentioned CW fiber laser with a 2-ns waveguide modulator, a 10-Hz Nd:YLF regenerative amplifier, and a single-shot three-stage Nd:glass amplification system, and it delievered an 1-J, 2-ns pulse at 1054-nm. The 2108-nm femtosecond OPA pulse mentioned above was stretched to 1.5-ns by pulse stretcher, then amplified to 80-mJ by OPCPA with a 22-mm-thick $LiNbO_3$ crystal and 1 $GW/cm^2$ pump intensity, and finally compressed into a 30-mJ, 300-fs pulse by pulse compressor. This 2108-nm femtosecond OPCPA laser was further converted to a 10-mJ femtosecond pulse at 1054-nm by SHG with a 4-mm-thick $\beta$-BBO crystal, which served as the real high-contrast test pulse in Fig. 3.

**SSCC device.** The key component of SSCC device is an adaptor that temporally magnifies test pulse and allows accommodation to photomultipler tube (PMT) and oscilloscope. Such adaptor consists of a correlation unit of noncollinear third-harmonic generation (THG) and a detection unit of fiber array and PMT. In the THG correlation unit applied in this work (Fig. 4), 80% of 1054-nm test pulse, reflected by a beam splitter, was frequency doubled to a 527-nm samping pulse by a 4-mm-thick β-BBO crystal ($\theta$=22.8°). Such sampling pulse, obtained by SHG, will be much more clean than test pulse. The left 20% of test pulse, after a half-wave plate, served as another incidence to THG correlation. The intersecting angle between test and sampling beam was 33.6°, and THG correlation adopted a type-I β-BBO crystal ($\theta$=61°) with a size of 38 ($x$) × 10 ($y$) × 2 (thickness) $mm^3$. Both the laser beams of test pulse and sampling pulse were nearly uniform in $x$ domain and covered the whole crystal. The noncollinear THG performed time-to-space mapping and induced a temporal window of 120 ps. The generated 351-nm correlation signal $A^{(2)}(x)$ with beam width $Wx$ = 38 mm was imaged onto a 12.5-mm-wide fiber-array by a cylindrical lens ($f$ = 100 mm). In SSCC detection unit, the fiber array consisted of 100 ultraviolet fibers (OPTRAN UVNS, CeramOptec) with an incremental length of 1 m, in which the 1st and the 100th fibers were 2.6 m and 101.6 m, respectively. The ultraviolet fiber with a core diamter of 105 μm had a high transmission (~10 dB/km) at 351 nm. The correlation signal $A^{(2)}(x)$ was coupled into the fiber array by a cylindrical lens with $f$ = 30 mm, and collected by a fiber bundle, which output a series of 5-ns-delayed temporal slices spanning 500 ns. Consequently, a serial instead of parallel detection can be used. In this work, PMT (H10721-113, Hamamatsu) was used as detector. To accommodate a high dynamic-range and ensure a linear response of PMT, a series of variable fiber attenuators should be applied to reduce the signal intensities in different fiber channels. In addition, a band-pass filter for 351-nm correlation signal was added between the fiber bundle and PMT to suppress the noise interference from two input pulses.

**SSCC modifications for dynamic-range test**. The SSCC device was slightly modified to evaluate the contrast measurement limit (Fig. 2). In the first modification, the spatially-uniform test beam was replaced by a synthesized high-contrast beam. An 1054-nm mirror with 40% reflectivity and 60% transmission combined together the femtoscond narrow beam and the CW wide beam. The transmitted femtosecond narrow beam had 25-μJ pulse energy, 150-fs duration, and 0.38 mm ($x$) ×1.8 mm ($y$) beam spot, rendering an intensity of 30 GW/cm$^2$. The reflected wide CW beam had 30-mW full power, and 38 mm ($x$) × 2.5 mm ($y$) beam spot, rendering a full intensity of 40 mW/cm$^2$. In the second modification, the 50-μJ femtosecond sampling laser at 527-nm was focused onto the correlation crystal, which produced a beam spot similar to the femtosecond narrow beam at 1054-nm, and a high intensity of 60 GW/cm$^2$. To achieve THG correlation function $A^{(2)}(x)$, this sampling laser was scanned in $x$ dimension by translation stages TS-2 and TS-3. Other parameters of the synthesized test beam and sampling beam have been introduced in the text. The third modification reduced noncollinear THG angle from 33.6° to 13.8° and $\beta$-BBO crystal orientation ($\theta$) from 61° to 36.5° in order to match the velocities of two incident waves. The correlation unit generated 9-μJ THG signal peak for the fiber pixel at $x$ = 0, corresponding to a conversion efficiency of 36% (pulse energy) and 12% (photon number) relative to 1054-nm narrow-beam femtosecond pulse. The total transmission from the correlation crystal to PMT was measured to be 25%.

**Single-photon detection in SSCC.** The input/output relationship of PMT is briefly introduced here. The PMT (H10721-113, Hamamatsu) applied in SSCC has an anode radiant sensitivity of $2.2\times10^5$ A/W. A photon at 351 nm has an energy of $5.66\times10^{-19}$ J, corresponding to $2.8\times10^{-10}$ W in the PMT characteristic time of 2 ns. According to the data listed above, a single photon of THG correlation signal triggers anode current of $6.16\times10^{-5}$ A, and further converts to a voltage of 3 mV with a coupling resistance of 50 Ω.

## Acknowledgements

This work was supported by grants from the National Natural Science Foundation of China (61727820).

## Author contributions

L. Q. and J. Ma. conceived the original ideas. J. M., P. Y., and X. O. designed and implemented the experiments. P. Y., J. W., and G. X. took part in the data analysis. L. Q. and J. Ma wrote the manuscript with inputs from all authors. L. Q. supervised the project.

## Conflict of interest

The authors declare that they have no conflict of interest.

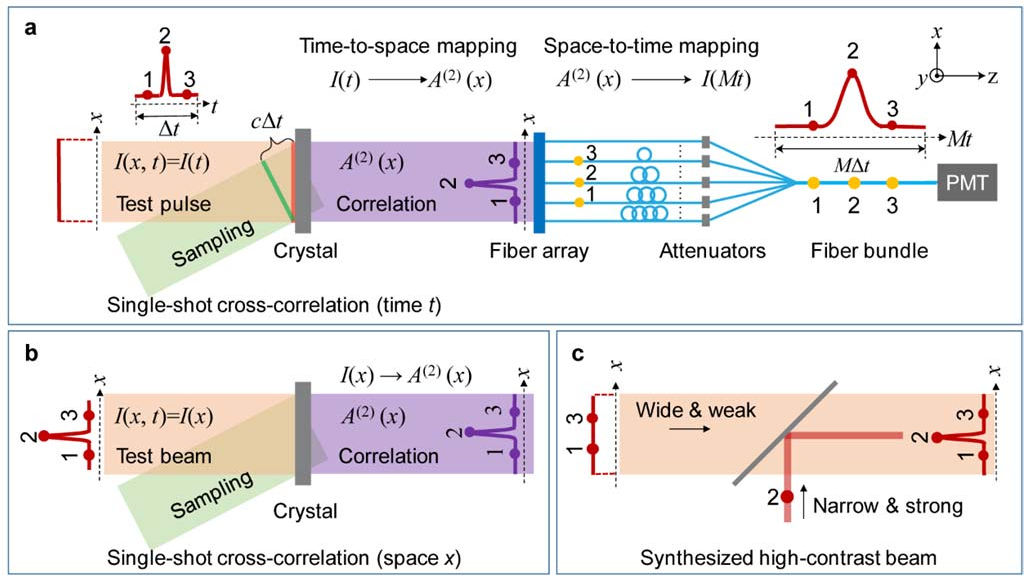

**Fig. 1 Temporal magnification and SSCC test strategy. a**, Test pulse *I(t)* represented by THG correlation function $A^{(2)}(x)$ with time-to-space mapping. Temporal magnification *M*, based on twice mapping between *t* and *x*, allows test pulse spanning *M*Δ*t* and accommodation to PMT and oscilloscope. **b**, Test beam *I(x)* represented by $A^{(2)}(x)$ without time-to-space mapping. **c**, Mimicking high-contrast pulse *I(t)* by $A^{(2)}(x)$ from a delicately synthesized beam *I(x)*.

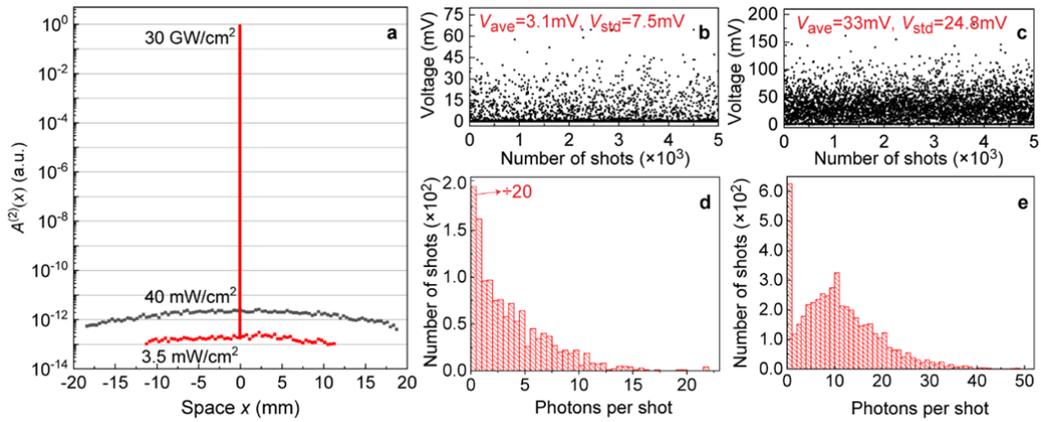

**Fig. 2 Dynamic-range test with a synthesized high-contrast beam. a,** Measured THG correlation function $A^{(2)}(x)$ at two CW intensities. Each data was averaged by 5000 shots with oscilloscope. **b**, **c** Recorded 5000-shot PMT photovolatages at $x$ = 0.38 mm and two CW intensities of 3.5 mW/cm² (middle) and 40 mW/cm² (right). **d**, **e** Histograms of the photon number detected per shot in correspondence to **b** and **c**, respectively.

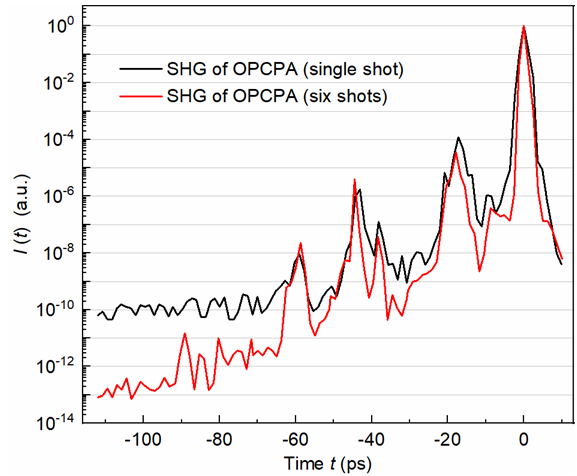

**Fig. 3. Two SSCC measurements for a real ultrashort pulse.** The black curve represents a full-window (120 ps) single-shot measurement at low intensity (10 GW/cm²), while the read curve consists of six equally-delayed small-window (24 ps) measurements at high intensity (50 GW/cm²).

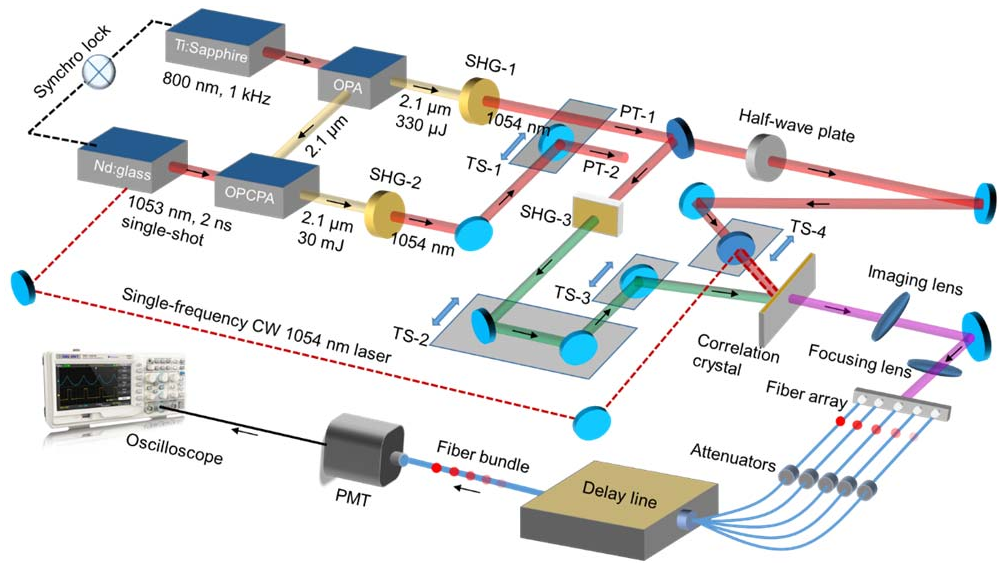

**Fig. 4. Schematic of experimental setup.** In brief, the setup consists of two laser sources (Methods for details) and SSCC device (Methods for details). 1-kHz, 2108-nm femotosecond OPA output acted either as the test laser after SHG (PT-1 beamline) applied in Fig. 2 or the seed to 2108-nm OPCPA pumped by 2-ns Nd:glass laser. SHG of OPCPA output (PT-2 beamline) was the test laser applied in Fig. 3. The single-frequency CW fiber laser in Nd:glass laser system was a part of the synthesized beam applied in Fig. 2. The roles of four translation stages: TS-1 and TS-4, selecting and adjusting the test laser; TS-2, adjusting the delay between test and sampling pulse; TS-3, spatially scanning the samping laser. SSCC device relied on a correlation unit of noncollinear THG and a detection unit of 100-pixel fiber-array. Finally, a series of 5-ns-delayed temporal slices spanning 500 ns, output from the fiber array (delay line), was collected by a fiber bundle, and detected by PMT and oscilloscope.